\documentclass[12pt,preprint2]{proto}

\usepackage{graphicx}

\newcommand\aastex{AAS\TeX}
\def\h2{H$_2$}
\def\-2e{$^{-2}$\ }
\def\-2{$^{-2}$}
\newcommand{\HI}{\mbox{\rm \ion{H}{1}}}

\newcommand{\x}[1]{\ensuremath{X_{\mathrm{#1}}}}
\newcommand{\ico}{\ensuremath{I_{\mathrm{CO}}}}
\newcommand{\cm}{\mbox{cm$^{-2}$}}
\renewcommand{\t}[1]{\mbox{#1}}

\newcommand{\ebv}{\ensuremath{E(B-V)}}
\newcommand{\ex}{\ensuremath{E(m_{\lambda_1}-m_{\lambda_2})}}
\newcommand{\ehk}{\ensuremath{E(H-K)}}
\newcommand{\ejh}{\ensuremath{E(J-H)}}
\newcommand{\av}{\ensuremath{\mbox{$A_{\rm V}$}}}
\renewcommand{\H}{\ensuremath{\t{H}}}
\newcommand{\xunits}{\mbox{cm$^{-2}$ (K km s$^{-1}$)$^{-1}$}}
\newcommand{\counits}{\mbox{K km s$^{-1}$}}

\newcommand{\nhunits}{\mbox{mag$^{-1}$ cm$^{-2}$}}
\newcommand{\msun}{\ensuremath{M_\odot}}

\begin{document}

   \title{Extinction in the Large Magellanic Cloud}

   \author{Nia Imara and Leo Blitz}
   \affil{Astronomy Department, University of California, Berkeley, CA 94720}
   \email{imaran@berkeley.edu}
  


   \begin{abstract}

We present an extinction map of the Large Magellanic Cloud (LMC), using 204,502 stars from the Two Micron All Sky Survey point source catalog.  We first use the NICE method to determine the reddening distribution, \ehk~and \ejh, which we compare to the HI distribution to find a near-infrared reddening law of $\ejh/\ehk=1.20\pm 0.04$.  A visual extinction map ($\sim 6^\circ\times 6^\circ$) of the LMC is created using the NICER method; at 4 arcmin resolution, a mean value of $\av=0.38$ mag is found.  We derive the LMC CO-to-H$_2$ conversion factor, $\x{LMC}$, independent of assumptions about the virialization of giant molecular clouds, by comparing the NICER extinction map with NANTEN $^{12}$CO observations.  In regions where $\av>1$ mag and $^{12}$CO emission is $\ge$ 2 \counits, we measure $\x{LMC}=9.3\pm 0.4\times 10^{20}~\xunits$.  In the same regions, the LMC contains a total molecular mass of $(4.5\pm 0.2)\times 10^7 ~\msun$.

   \end{abstract}

%


\section{Introduction}

Studying the properties of giant molecular clouds (GMCs), the sites of most star formation, in a variety of extragalactic environments provides insight into stellar evolution as a whole.  The Large Magellanic Cloud (LMC) is a good laboratory for this type of study because of its proximity ($D\sim 50$ kpc) and its nearly face-on orientation in the sky. 

Determining the amount and distribution of molecular hydrogen (H$_2$), the dominant molecule ($\sim 99.99\%$ by number) in GMCs, is useful for the study of the initial conditions of star formation.  Yet due to the lack of a permanent electric dipole moment, H$_2$ is undetectable at the low temperatures ($\sim 10$ K) of molecular clouds, and so a variety of other tracers have been used to identify them and to study their density distribution.  Two common methods are star counts and color excess, which trace molecular gas by measuring the dust distribution of GMCs.  The spatial location of gas and dust are highly correlated, as the well-known gas-to-dust relationship attests, because H$_2$ forms most efficiently in dense gas where dust grains provide surfaces onto which \HI~can collide and be converted into molecular form.   The star counts method measures extinction by comparing the number of stars seen behind a dark cloud along a given light of sight with the number in a control field off the clouds and assumed to be dust-free.  The color excess method measures reddening by exploiting the decrease in the extinction with increasing wavelength.

Since the 1960s, a number of authors have used the color excess method to map the reddening and inferred dust distribution of the LMC, (e.g.,  Feast et al. 1960; Isserstedt 1975; Grieve \& Madore 1986).  But a combination of questionable assumptions---for instance, assuming Galactic intrinsic colors, small data sets, limited spatial coverage, and not accounting for the Galactic foreground contribution to reddening---resulted in significant uncertainties.  

Advances in recent years have made it possible to possible to bypass most of these problems. Namely, the advent of infrared (IR) array detectors, such as those used in the Two Micron All-Sky Survey (2MASS), have made possible the simultaneous observations of hundreds of sources in multiple wavelength bands.  These detectors have the further advantage that IR light suffers much less extinction than the visible and ultraviolet, so molecular clouds can be probed more deeply.  In the Galaxy, for instance, $A_{\rm K}\approx 0.1\av$.  The Near Infrared Color Excess (NICE) technique developed by Lada et al. \cite{lada} exploits the advantages offered by these large-scale arrays and has proved to be a powerful way of mapping extinction through molecular clouds.  In this method, infrared color excess measurements are made by inferring the intrinsic color, typically $(H-K)$, of target stars from an unreddened control field. The NICER (NICE Revised) method of Lombardi \& Alves (2001) generalizes the former by using information from all available independent colors---for instance, $(J-H)$ and $(H-K)$---to map extinction.

Radio observations of interstellar carbon monoxide (CO) have also been used to trace H$_2$.  Cohen et al. \cite{cohen} were the first to fully survey the LMC in the $J=1\rightarrow 0$ rotational transition of CO.  Yet, at their limited resolution ($8.'8$), they were able to resolve only the largest cloud complexes ($\sim 140$ pc).  More recently, Mizuno et al. \cite{mizuno} completed a comprehensive high-resolution ($\sim 40$ pc) CO survey using the 4 m NANTEN telescope.  Both groups of authors made calculations of the CO-to-H$_2$ conversion factor, or the ``$X$-factor,'' of the LMC.  

An outstanding question concerning GMCs is how metallicity affects the $X$-factor.  This quantity is of particular interest in the LMC since the galaxy has a lower metallicty ($\sim 1/4$; Dufour 1984) and a higher gas-to-dust ratio ($\sim 1.7$; Gordon et al. 2003) than the MW.  It is conjectured that the relatively less-abundant CO in the LMC is more supceptible to photo-dissociation by UV radiation, which can penetrate more deeply into the less-dusty clouds (e.g., McKee 1989).  This argues for $\x{LMC}>\x{MW}$.  

Yet there are a number of problems with determining the $X$-factor soley from CO observations.  Potentially unrealistic assumptions  regarding the physical properties of the molecular clouds---for instance, that they are virialized---must be made, since there is not an independent measure of $N(\H_2)$.

It is the main goal of this paper to provide such an independent measure of the molecular content in the LMC, one that makes no assumption about the structure, dynamics, or virialization of GMCs. We do this by first creating an extinction map from 2MASS data, in one of the first applications of the NICE(R) techniques to an extragalactic system.  A number of other useful results are obtained in the process, including the NIR extinction law, the extinction distribution, and the $X$-factor of the LMC.  In \S~2 we describe data selection and the other considerations that go into determining extragalactic reddening.  The results, including the extinction map and $X$-factor, are presented in \S~3.  Conclusions are presented in \S~4.


\section{Data Selection}

\subsection{2MASS Data}

Our reddening maps of the LMC are derived using $JHK$ near-IR photometry obtained from the 2MASS Point Source Catalog\footnote{http://irsa.ipac.caltech.edu/applications/Gator/}.  For unconfused sources with Galactic latitude $|b|>10^\circ$, the limiting magnitudes of the 2MASS catalog at the 10-$\sigma$ level are 15.8, 15.1, and 14.3 mag in $J~(1.24~\mu$m), $H~(1.66~\mu$m), and $K_{\rm S}~(2.16~\mu$m), respectively.  Possible contaminants, such as image artifacts caused by bright stars, confused sources, solar system objects, and other artifacts are eliminated\footnote{Explanatory Supplement to the 2MASS All-Sky Data Release at http://www.ipac.caltech.edu/2mass/releases/ allsky/doc/explsup.html}. Only data having photometric uncertainty $\le 0.10$ mag, corresponding to $S/N\ge 10$, are selected.  Our data cover the central $6^\circ\times 6^\circ$ region of the LMC, centered on $(\alpha,\delta)=(5^{\rm h}20^{\rm m},-69^\circ$).  This choice of range was based on an examination of the Mizuno et al. \cite{mizuno} CO map and the Staveley-Smith et al. \cite{s-smith} \HI~ map, which reveals the extent of molecular and \HI~emission, respectively.  This initial sample contains 329,702 sources.

The 2MASS resolution, $2^{\prime\prime}$, corresponds to a physical size of $\approx 0.5$ pc at 50 kpc, and so individual LMC stars should be well-resolved.  Nikolaev \& Weinberg \cite{NW00}, treating a similar set of LMC data drawn from 2MASS, show that confusion is not a problem at this resolution.

Our preliminary data set contains not only LMC field stars but Galactic foreground point sources.  In order to get a quantitative estimate of the MW foreground stellar number density, we obtain 2MASS data for a reference field in the same range of Galactic latitude as the LMC field but removed in Galactic longitude by $-30^\circ$.  The angular number densities of the LMC field (prior to data reduction) and MW field are $4.65\times 10^3$ deg$^{-2}$ and $1.68\times 10^3$ deg$^{-2}$, respectively.  Not accounting for the foreground would dilute the measured extinction (\av) and increase the uncertainty, especially in regions of high \av.  In what follows we use the color-color diagram, in conjunction with Nikolaev \& Weinberg's \cite{NW00} analysis of the LMC color-magnitude diagram, to correct for this. 

\subsection{Data Reduction}\label{sec:data_reduct}

To derive the LMC gas-to-dust ratio (see \S~\ref{sec:extlaw}) and $X$-factor (\S~\ref{sec:xfactor}), we generate color excess and visual extinction maps for comparison with the galaxy's $N(\HI)$ and CO distributions, respectively.  We use Lada et al.'s (1994) NICE  method to generate the color excess maps, $E(J-H)$ and $E(H-K)$.  For the extinction  maps, we use the NICER method of Lombardi \& Alves (2001), which generalizes Lada's formulation for multi-band observations. In these techniques, the intrinsic color of all stars observed toward a target clouds is taken to be the mean color of a nearby, unreddened control field. An extinction map is generated by averaging the irregularly sampled data into pixels, across each of which the extinction is assumed to be uniform.  Application of these techniques, originally applied to Galactic clouds, assumes that  all stars observed toward the cloud (i.e., the field stars) are background stars; and the field and control stars are homogeneous (i.e., drawn from the same parent population). Extending these techniques to extragalactic sources raises special concerns as to how well these conditions can be fulfilled.  

Before creating the extinction map, a number of reductions and corrections are performed on the raw data, with particular care taken to address those assumptions just mentioned.

\subsubsection{Completeness}

\begin{figure}[ht!]
\epsscale{1.0}
\plotone{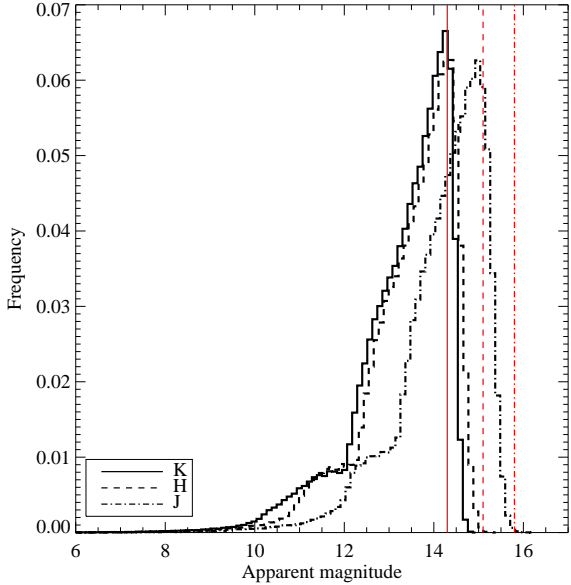}
\caption{Luminosity functions of our sample.  The observed completeness limits for $J$, $H$, and $K$, were estimated 15.0, 14.4, and 14.3 mag, respectively.  The 2MASS completeness limits (15.8, 15.1, and 14.3 mag) are indicated with vertical lines. \label{complete}}
\end{figure}

The observed completeness limits toward the LMC are brighter than the nominal completeness limits stated in the 2MASS catalog (Figure~\ref{complete}).  Towards regions of higher extinction, intrinsically fainter stars will escape detection before intrinsically brighter stars, so the overall extinction would be underestimated if all stars were included in the measurements (Oestreicher \& Schmidt-Kaler 1996).  Because only the minimally extincted faint stars would be able to be measured, including them in our sample would drive down the overall mean extinction.  However, in dealing with apparent, not absolute, magnitudes, the relative faintness of a given star may be due to extinction or its intrinsic brightness.  But statistically, we do not expect a particular population (intrinsically faint {\em or} bright) to be more likely to lie behind molecular clouds.  With our large data set, then, it is reasonable to assume that most observed faint stars (i.e., those below the observed completeness limit) would bias measurements to lower values of extinction if included.  Thus, based on our completeness estimates, we used only stars with brightness greater than $J<15.0$, $H<14.4$, and $K<14.3$ mag for subsequent analaysis.  This and the following data selections are summarized in Table~\ref{tab:data}.

We tested these assumptions in several ways.  First, we confirmed that the use of a constant magnitude limit in each band is appropriate, i.e., that it does not depend on column density.  We calculated the limits for different reddening intervals and found that the values stated above remained the same.  Second, we measured the extinction for those stars falling outside our stated limits (i.e., those fainter than $J>15.0$, $H>14.4$, and $K>14.3$ mag).  Their measured mean reddening of -0.44 mag is unphysical and their distribution is consistant with an error population.  Indeed, when we went ahead and created a map including these stars, (i.e., without having imposed a completeness limit), the global mean extinction was lowered significantly (by nearly a factor of 2).  Finally, after having created the extinction map only using stars brighter than the completeness estimates, we found that the remaining fainter stars have overall lower reddenings consistant with an error population (i.e., a noise population with a Gaussian distribution centered around zero).

\subsubsection{Foreground and non-interstellar reddening}

The Galactic foreground contaminates the 2MASS LMC field in two ways: with \HI~gas and point sources.  Within the LMC itself there is foreground contamination, as well as reddening of non-interstellar origin, including that caused by protostars and obscured aymptotic giant branch (AGB) stars.  

Galactic \HI~foreground extinction is not negligible but previous authors, usually for lack of data, adopted uniform foreground reddening in their derivation of LMC extinction maps.  In the Parkes multibeam \HI~ survey of the LMC, Staveley-Smith et al. (2003) found a mean Galactic reddening of $\langle E(B-V)\rangle=0.06$ mag over the LMC disk, varying from 0.01 to 0.14 mag.  Staveley-Smith generously provided us with their \HI~data, which we used to correct for Galactic reddening for stars in the LMC field.  For every 2MASS source, foreground extinction in each band was calculated at that position using the MW extinction law of Schlegel, Finkbeiner \& Davis \cite{sfd}~($A_{\rm J}/\av=0.276$, $A_{\rm H}/\av=0.176$, $A_{\rm K}/\av=0.112$), and the MW gas-to-dust ratio of Bohlin et al. \cite{bohlin}~($\beta\equiv N(\HI)/\av=1.87\times 10^{21}$ mag$^{-1}$ cm$^{-2}$).  These values were then subtracted from the observed 2MASS magnitudes:

\begin{eqnarray}
J_{\rm corrected}=J_{\rm observed}-\frac{0.276}{\beta}  N(\HI)_{\rm foreground}   ,\\
H_{\rm corrected}=H_{\rm observed}-\frac{0.176}{\beta}  N(\HI)_{\rm foreground},\\
K_{\rm corrected}=K_{\rm observed}-\frac{0.112}{\beta}  N(\HI)_{\rm foreground} ,
\end{eqnarray}

While Galactic \HI~is easily identified given its distinct velocity compared to LMC gas, the correction for Galactic point source contamination is more complicated.  Furthermore, we wish to eliminate from our sample objects in the LMC whose reddening is circumstellar in origin.  The color-magnitude diagram (CMD) is a useful tool to correct for both effects.  Analysis of the CMD allows us to eliminate sources, on a {\em probablistic} basis, that are unlikely to have undergone reddening from LMC interstellar dust.

\begin{figure}[ht!]
\epsscale{1.}
\plotone{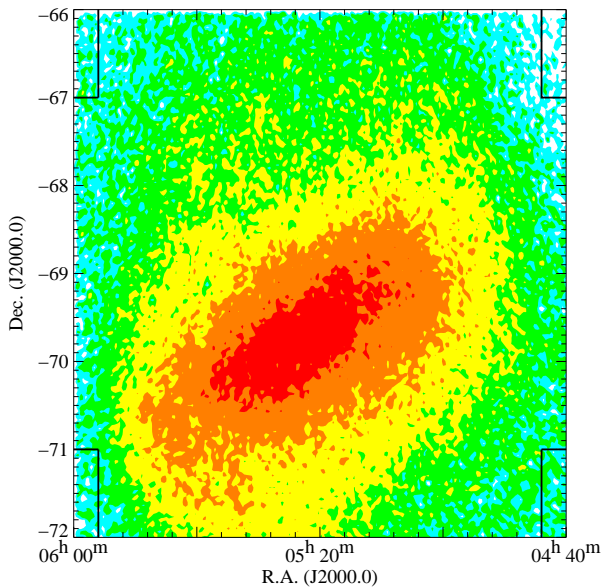}
\caption{Spatial distribution of the selected sources, in pixels of $4'\times 4'$, with average surface density of $\Sigma_{\rm{LMC}}=4.76\times 10^3$ stars deg$^{-2}$.  Counterclockwise starting from the northeast, the outlined control regions have an average of $\Sigma_{\rm{cont.}}=1.48$, 1.57, 1.98, and $1.02 \times 10^3$ stars deg$^{-2}$.  The color levels cyan, green, yellow, orange, and red correspond to  0.90, 1.80, 3.60, 7.20, and 14.40 $\times 10^3$ stars deg$^{-2}$, respectively. \label{numden}}
\end{figure}

\begin{figure}[ht!]
\epsscale{1.0}
\plotone{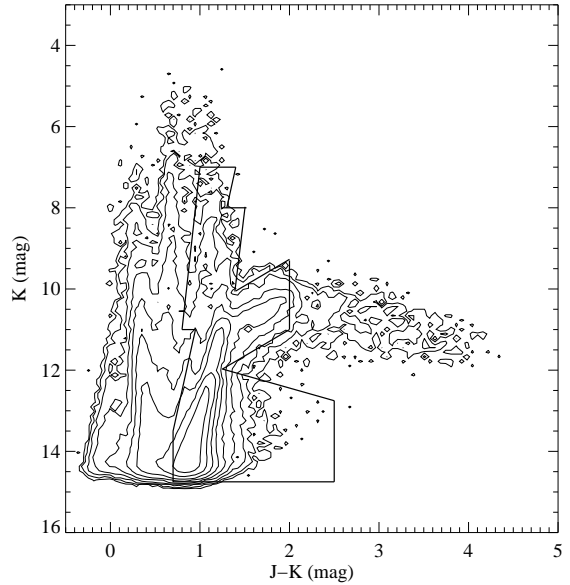}
\caption{Color-magnitude diagram of the LMC field.  The contours are logarithmically spaced, from 2 to 3.4, by 0.4. We selected data for the extinction maps from the enclosed region, which corresponds to regions E, F, G, H, J, L and part of D of Nikolaev \& Weinberg (2000). \label{jkk}}
\end{figure}

\begin{figure}[ht!]
\epsscale{1.0}
\plotone{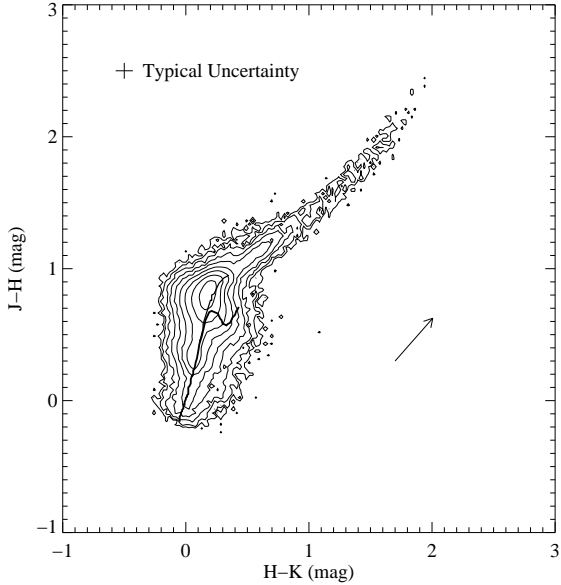}
\caption{Color-color diagram of the LMC field before data reduction.  The contours are logarithmically spaced, from 2 to 3.8, by 0.4.  Overplotted are the color sequences for Galactic dwarfs and giants from Koornneef (1983).  The reddening vector, based on relations from Koornneef (1982), is drawn for $\ebv=1.0$ mag.   \label{jhk1}}
\end{figure}

\begin{figure}[ht]
\epsscale{1.0}
\plotone{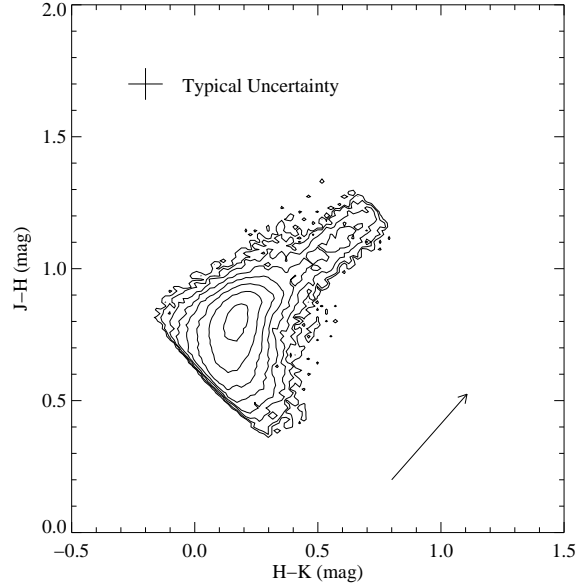}
\caption{Color-color diagram of the 204,502 stars in the LMC field after data reduction.  The contours and reddening vector are as in Figure~\ref{jhk1}.   \label{jhk2}}
\end{figure}

\begin{table}[ht]
\begin{center}
\caption{Data selection summary.\label{tab:data}}
\begin{tabular}{ll}
\tableline\tableline 
Criteria & Number of \\
         & remaining stars  \\
\tableline
2MASS selection   & 329,702  \\
Completeness cutoff\tablenotemark{a}	& 258,621  \\
Eliminate MW foreground\tablenotemark{b}& 204,502  \\
\tableline
\end{tabular}
\tablenotetext{a}{Completeness limits (see Figure~\ref{complete}): $J<15.0$ , $H<14.4$,\\
 $K<14.3$ mag.}
\tablenotetext{b}{Based on Nikolaev \& Weinberg \cite{NW00} regions.}
\end{center}
\end{table}


One can use the CMD ($K$ vs. $J-K$) to identify distinct stellar populations (Figure~\ref{jkk}).  This is the main focus of Nikolaev \& Weinberg \cite{NW00} who, by matching features of the 2MASS CMD to colors of known populations from the literature, discuss which populations are the greatest contributors to the different regions.  We use their regions E, F, G, H, J, K, L, and part of D, to create the reddening maps.  These areas correspond to LMC populations, primarily of the giant branch, with insignificant Galactic foreground contamination.  We eliminate regions A, B, C, and I, which contain roughly 15, 80, 80, and 55 percent foreground, respectively.  Region K, which is free of foreground contamination, but consists of dusty asymptotic giant branch stars whose large $J-K$ colors are due to circumstellar reddening, is also eliminated.  We confirmed that these highly reddened stars are not preferentially located in regions of high molecular gas by plotting their spatial locations.  We also performed the reverse operation.  That is, based on the CO map of Mizuno et al. \cite{mizuno}, we located points on the color-color diagram and CMD that fall within the contours of known molecular clouds.  In all, there are only 74 stars in the extended ``finger'' of the color-color diagram corresponding to molecular clouds in the Mizuno et al. \cite{mizuno} map (see Figure~\ref{jhk1}).  After these selections, we are left with 204,502 sources, corresponding to an average number density of $4.76\times 10^3$ deg$^{-2}$.  Figures~\ref{jhk1} and~\ref{jhk2} show the color-color diagrams of the LMC field before and after the data reductions presented in this section. 
 
A difficult problem is correcting for foreground contamination within the LMC.  The color excess method assumes that stars observed in the line-of-sight to a cloud are background stars. Extinction measurements will be diluted wherever foreground stars are included in the calculations. In addition, as discussed and modeled by Lada et al. (1994), the decrease in measured extinctions due to foreground stars will have a particularly large impact in regions of high extinction, resulting in an increase in dispersion with \av. { We tried to minimize the effect of foreground in the extinction map by sigma-clipping---that is, by removing sources in a given pixel with anomolous values of \av.  This was one of the motivations behind the creation of the NICER technique (Lombardi \& Alves 2001).  Described comprehensively by these authors, the sigma-clipping solution is justifiable when the scatter of intrinsic stellar colors is small or when the column density of the GMCs in question is sufficiently high.  As we show in the following subsection, the first condition is not met in our case.  Neither do we have any {\em a priori} reason to sigma-clip based on the latter condition.

In any case, to explore the effects of this method on our data set, we did generate a number of sigma-clipped maps.  Ultimately, we found no advantages in doing so and used a simple weighted mean for our final maps.

\begin{figure}[ht]
\epsscale{1.0}
\plotone{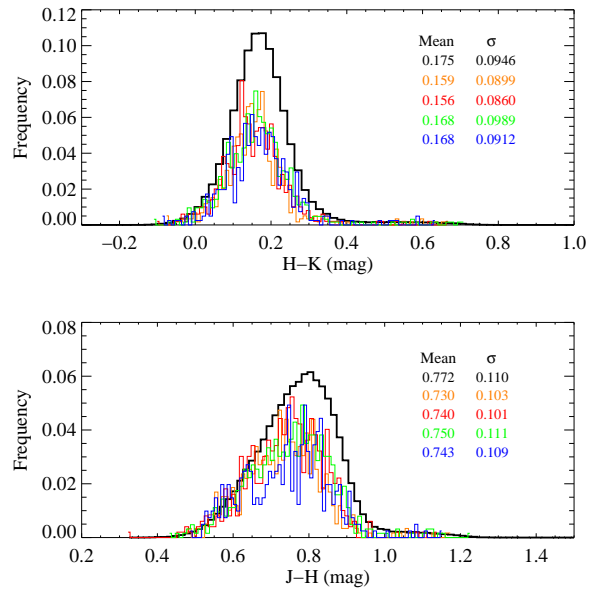}
\caption{The reddening distrutions of the LMC field (heavy black line) and control (colored lines) groups.  The orange, red, green, and blue lines correspond to data from the northeast, southeast, southwest, and northwest corners of the LMC, as shown in Figure~\ref{numden}. \label{control}}
\end{figure}

\subsubsection{Homogeneity of control and field stars}

Success with the NICE(R) technique relies on the mean color of the control group being characteristic of the intrinsic color of the field stars. Recall, the reddening of a given star is the difference between the observed color and the intrinsic color {\em of a particular spectral type}.  Thus, extending the NICE(R) method to an entire extragalactic field raises another concern as to whether the fraction of stars of a particular spectral type varies significantly across the field.  In other words, we are concerned with the homogeneity of the field, especially the control group.  Based on the selection criteria in the previous section, we are confident that our field consists primarily of giant branch stars, though a variety of stages in evolution are represented (e.g., helium and hydrogen shell burning phases, oxygen- and carbon-rich phases; Nikolaev \& Weinberg 2000).  As for the control group, we would like the dispersion around the mean control color to be significantly less than the observed range of field star colors, since the dispersion determines the 1-$\sigma$ confidence level for extinction measurements (Alves et al. 1998). 

We examined the \HI~and CO maps of the LMC for minimally reddened regions and inspected control groups in each of the four ``corners'' of the galaxy image (Figure~\ref{numden}).  Not surprisingly, the regions do not have narrow color distributions.    The control groups are well-behaved in that they are statistically equivalent---the $(H-K)$ colors to two significant figures and the $(J-H)$ colors to one significant figure.  Yet, as Figure~\ref{control} shows, the width of the control distributions relative to that of the field stars forecasts low signal-to-noise in extinction measurements.  The nature of the uncertainties is explored in detail in \S~\ref{sec:nicer} where we derive the extinction maps.  We are somewhat consoled, though, since the large number of stars in our data set will act to minimize this problem and because our main interests here are in global properties of the galaxy.  We proceed by taking the average of the mean control group colors as the intrinsic colors of the field:

\begin{eqnarray}
(H-K)_{\rm{intrinsic}} &\equiv &\langle (H-K)\rangle_{\rm{control}} \nonumber \\
	&=& 0.16\pm 0.09~\t{mag},\label{eq:control1}\\
(J-H)_{\rm{intrinsic}} &\equiv& \langle (J-H)\rangle_{\rm{control}} \nonumber \\
	&=& 0.74\pm 0.11~\t{mag}.\label{eq:control2}
\end{eqnarray}

\noindent By comparison, the mean colors of the entire data set are $(H-K)=0.175\pm0.0946$ and $(J-H)=0.772\pm 0.110$ mag (see Figure~\ref{control}). 
 Again, this and the previous selection criteria are summarized in Table~\ref{tab:data}.


\section{Results}\label{sec:results}
\subsection{NIR Extinction Law}\label{sec:extlaw}

\begin{figure}[ht!]
\epsscale{.95}
\plotone{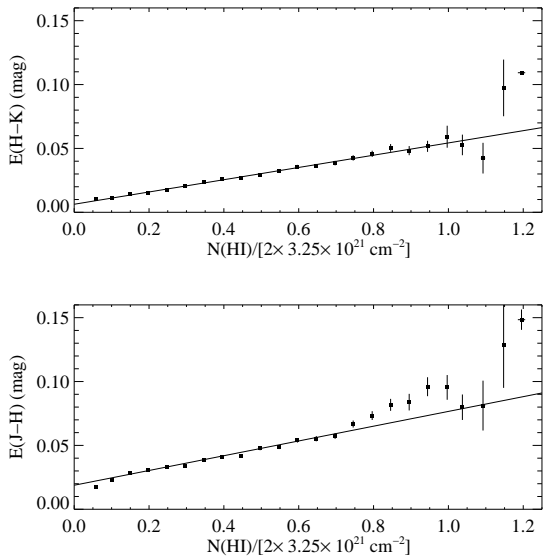}
\caption{LMC reddening as a function of hydrogen column, measured in regions where $\ico<1~\counits$.  Error bars represent the deviation around the mean of color excesses inside the bin ($\Delta N(\HI)/[2.\times 3.25\times 10^{21}$cm$^{-2}$=0.05]).\label{gasdust}}
\end{figure}

\begin{figure}[ht!]
\epsscale{.95}
\plotone{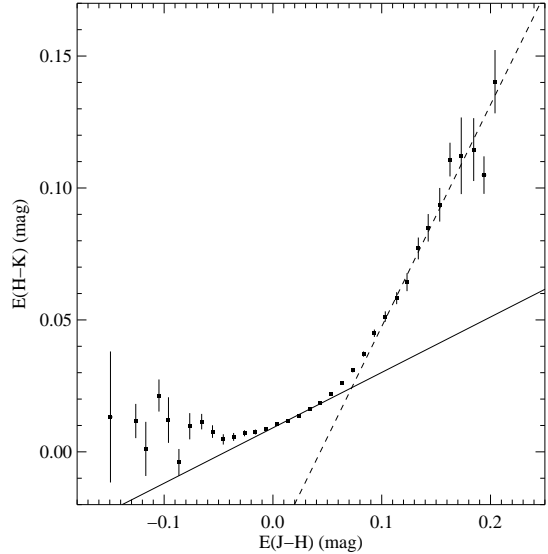}
\caption{The reddening law measured in the same regions as Figure~\ref{gasdust}.  The $(H-K)$ color excess is plotted as a function of $(J-H)$ color excess.  Error bars represent the deviation around the mean of \ehk~inside the bin $\Delta\ejh=0.01$ mag.  The data are plotted as a function of \ejh, as opposed to \ehk~ as is often the case, because in this circumstance \ejh~ has a higher overall signal-to-noise.\label{rlaw}}
\end{figure}

The NICE(R) methods require knowledge of the NIR extinction law and assume that this relationship is linear.  The extinction law may be described by extinction coefficients, $k_i=\av/\ex$, which relate the extinction to the color excess.  Knowing $k_i$ enables us to convert color excess to \av, which is proportional to $N(\H_{\rm total})$ and, hence, to $N(\H_2)$. To determine the scaling we must assume a value of the LMC gas-to-dust ratio.  Thus, we plot  \ex~versus $N(\HI)$, and compare this relationship with the LMC gas-to-dust ratio estimated by Gordon et al. (2003) (Figure~\ref{gasdust}).  Their value of $N$(\HI)$/\av=(3.25\pm 0.28)\times 10^{21}$ mag$^{-1}$ cm$^{-2}$ is a weighted average of all 24 archival gas-to-dust ratio measurements.  

\ehk~and \ejh~for individual stars were determined by subtracting the intrinsic colors (Equations~\ref{eq:control1} and~\ref{eq:control2}, respectively) from the observed 2MASS colors.  Color excess maps were generated by binning the data into $4'$ pixel grids with $2'$ spacing.  The color excess at a given location was then given by the average, weighted by the uncertainties, of stars within that pixel.  Next, these maps were compared to the $N(\HI)$ map on a pixel-by-pixel basis.


\begin{figure}[ht!]
\epsscale{.95}
\plotone{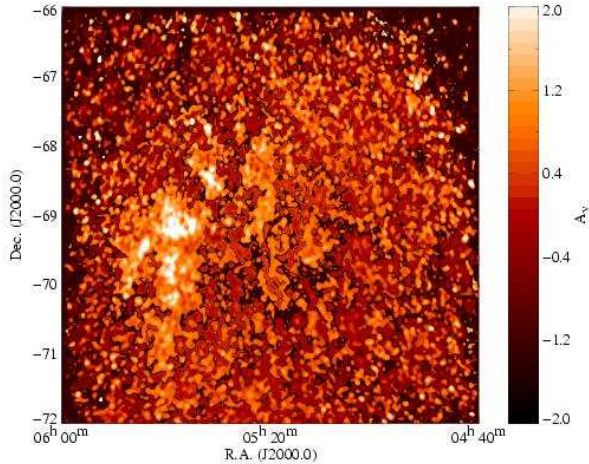}
\caption{Sigma-clipped NICER extinction map of the LMC at a resolution of 4 arcmin.\label{avmap1}}
\end{figure}

In Figure~\ref{gasdust}, the data are averaged in bins of equal $N(\HI)$ and fitted for the linear relation
\begin{equation}\label{eq:extlaw}
\ex=m_i\av(\HI)+b,
\end{equation}

\noindent where $m_i=k_i^{-1}$ and $\av(\HI)=N(\HI)/(2\times 3.25\times 10^{21}~\nhunits)$ is the extinction due to \HI.   The \HI~column is divided by 2 since, on average, extinction along a given line-of-sight is caused by half the dust layer.  In other words, only half the total column, on average, contributes to the measured reddening.  The data are plotted in bins of $\Delta\av(\HI)=0.05$ mag, and weighted least-squared fits were done over the range $\av(\HI)=[0.1,1]$ mag.   So as to get the best possible correlation between color excess and atomic hydrogen we masked out regions with significant molecular gas, defined as areas with CO emission exceeding $\ico\cong 1$ \counits.  This value is based on an examination of the Mizuno et al. \cite{mizuno} CO map, in which they define GMC complexes as reions exceeding 3 \counits.  The extinction coefficients $k_i$, which came from the inverse of the best fit slopes, were found to be $20.83\pm 0.52$ and $17.30\pm 0.46$ for \ehk~and \ejh, respectively.  Dividing these results gives a ratio of $\ejh/\ehk=1.20\pm0.04$, which is consistent with the Koornneef (1982) reddening law, which is $\ejh/\ehk=1.06$ when converted to the 2MASS photometric system (Carpenter 2001).

In principle, one should be able to determine the NIR reddening law by comparing \ehk~and \ejh~directly. And if the color excesses are linearly dependent, the ratio of extinctions in two different bands is constant, a key assumption of the NICE(R) techniques. We found that this direct comparison was not a robust way of measuring the reddening law, in part, because of the lack of a linear relation for low values of $(H-K)$ and $(J-H)$---a result of our data selection (see Figure~\ref{jhk2}).   When we examined the reddening law in the same regions as the gas-to-dust ratio analysis (i.e., where $\ico< 1~\counits$), we found that the data could be fit by two different slopes.  Figure~\ref{rlaw} shows that the majority of stars with little to no reddening are fit by a shallower slope than more highly reddened stars.  Below $\ejh\approx 0.05$ mag, the data are consistent with a noise population and there is no robust linear fit.  Above $\ejh\approx 0.1$ mag, however, the data are well approximated by the linear relationship $\ehk=(0.84\pm0.06)\ejh-(0.037\pm0.007)$.  The reciprocal of the slope, 1.19, is in good agreement with the $\ejh/\ehk$ ratio of 1.20 we derive via the gas-to-dust ratio measurements and with the Koornneef (1982) reddening law.  We also repeated this analysis for the entire reddening maps, i.e., not restricted to regions of low molecular gas.  The same result was observed: more highly reddened stars were fit by a linear relationship in agreement with a reddening law with a slope of $\approx 1.20$. 

Finally, we note that the lack of linearity in the reddening law over the entire range of values is not due to poor photometry. When we repeated the above analysis using stars with increasingly higher $S/N$ ratios, we achieved similar results: stars reddened above $\ejh\approx 0.1$ mag were well-fit by a reddening law of $\approx 1.20$, while stars with lower reddening were consistent with noise.

\begin{figure}[ht!]
\epsscale{.95}
\plotone{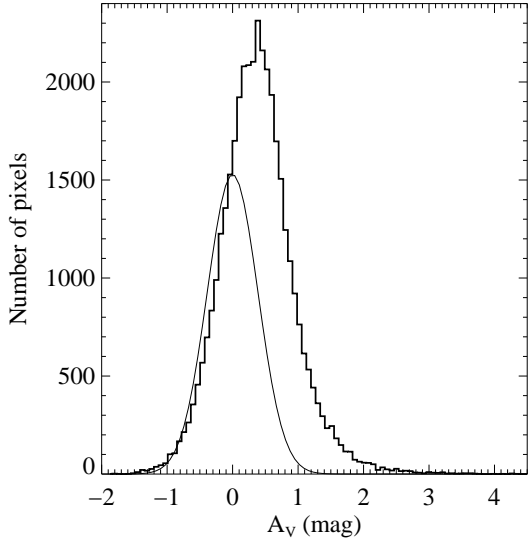}
\caption{The pixel extinction distributions of the  the sigma-clipped map (Figure~\ref{avmap1}).  Indicating a population of zero-reddening stars is a Gaussian with $\sigma$ equal to the mean of the propagated measurment errors of the data.\label{avhist}}
\end{figure}

\subsection{NICER Exinction Map}\label{sec:nicer}

Using the \ex-\av~ratios derived above, an extinction map, shown in Figure~\ref{avmap1}, was constructed using the NICER method (see Lombardi \& Alves 2001 for a detailed description).  The intrinsic colors and color dispersions, used to calculate the color excess for individual stars, were derived from the control fields (Equations~\ref{eq:control1} and~\ref{eq:control2}).  Visual extinctions were then calculated for each star, and the data were binned onto $4^\prime$ pixel grids with $2^\prime$ spacing.  At this resolution, we can resolve structure down to $\approx 58$ pc.  Pixels were required to contain at least three stars for subsequent calculations; those with $N<3$ were given a ``not-a-number'' value and interpolated over in the extinction map.  Out of a total of 44,590 ($245\times 182$), there are only 6,089 such pixels, primarily in the periphery of the map where the number density is low.

\begin{figure*}[ht!]
\epsscale{1.9}
\plotone{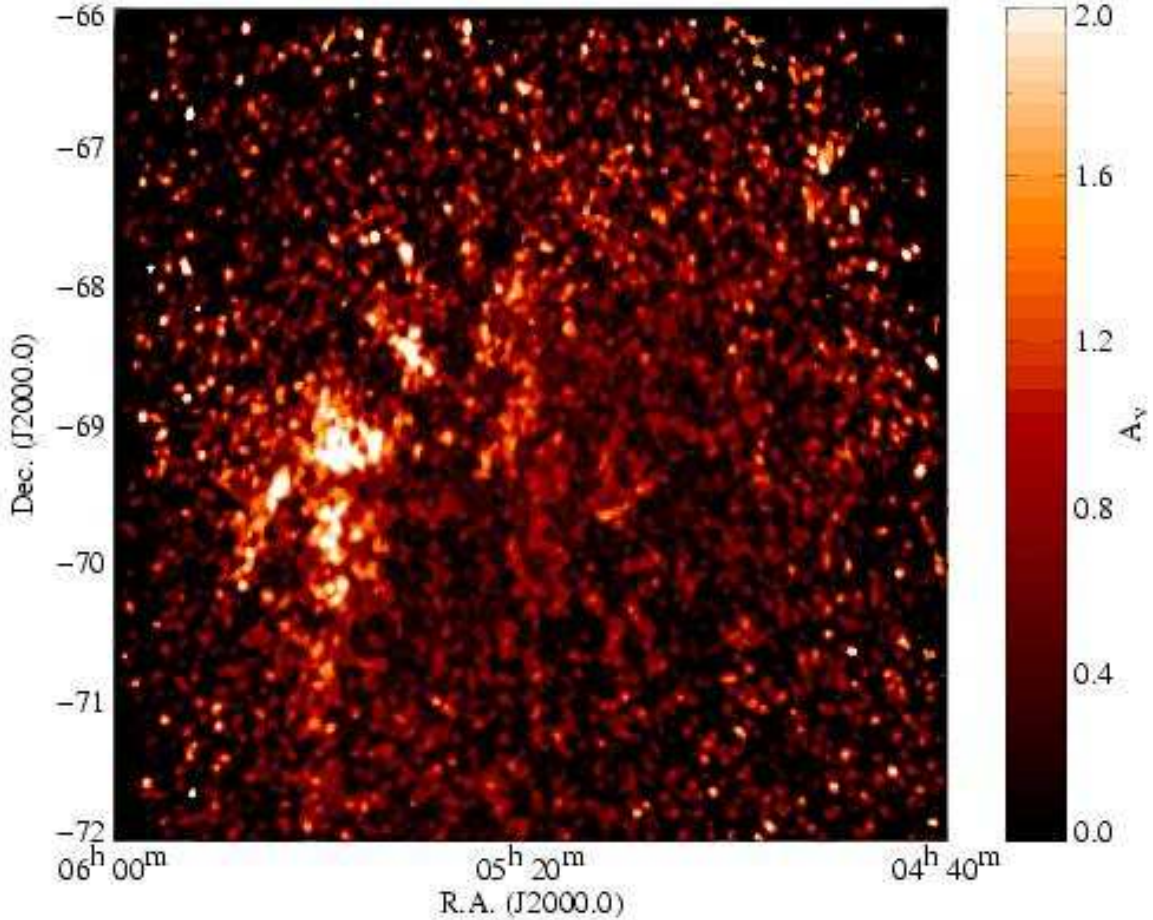}
\caption{Sigma-clipped NICER extinction map of the LMC at a resolution of 4 arcmin.  This map shows physically allowable extinctions, i.e., $\av\ge 0$ mag.\label{avmap0}}
\end{figure*}

At our resolution, we found a mean extinction (\av) of 0.38 mag and standard deviation of 0.57 mag. The distribution of measured values is displayed by the histogram in Figure~\ref{avhist}. ``Negative'' extinction pixels in Figure \ref{avmap1} correspond to regions that are bluer than the average intrinsic colors or are due to errors in the measurements.  For clarity, a map showing physical extinctions $\av\ge 0$ is shown in Figure \ref{avmap0}.  The most conspicuous feature of the map, which shows extinction due to both the atomic and molecular ISM of the LMC, is 30 Doradus at $\alpha\sim 5^{\rm h}40^{\rm m}$ and $\delta\sim -69^\circ$ to $\delta\sim -71^\circ$ (J2000.0).  Other filamentary and arc-like structures are seen, for instance, at $\alpha\sim 5^{\rm h}35^{\rm m}$ and $\delta\sim -68^\circ 20'$ to $-69^\circ 20'$.  The ``CO arc'' identified by Mizuno et al. \cite{mizuno} is also present in our map, extending southeast of 30 Doradus.  Extending southwest from  $(\alpha,\delta)\sim(5^{\rm h},-66^\circ)$ to $(4^{\rm h}40^{\rm m},-68^\circ 30')$ is a loose ``string'' of highly extincted clumps, some of which are associated with \HI~peaks.  But there are no peaks in either the \HI~or CO maps corresponding to the rather large ($\sim 175\times 90$ pc) clump at $(4^{\rm h}40^{\rm m},-68^\circ 30')$.  We do not find distinct, high-\av~features located toward the bar at $(\alpha,\delta)\sim( 5^{\rm h}20^{\rm m},-70^\circ)$.

\subsubsection{Uncertainties and noise}

The average extinction of a given pixel is estimated by the weighted mean of all stars falling within it: 

\begin{equation}\label{eq:av}
\langle\av\rangle=\frac{\sum_{i=1}^N W_i \av_{,i}}{\sum_{i=1}^N W_i},
\end{equation}

\noindent where the weight for the $i$th star is given by $W_i=1/\sigma_i^2$, and $\sigma_i^2$ is the variance due to the uncertainties in photometry and intrinsic star colors, $\sigma_i^2=\sigma_{\rm{phot},\it i}^2+\sigma_{\rm{color},\it i}^2$.  The error in the mean, or the observed variance, for the pixel is given by

\begin{equation}\label{eq:sig}
\sigma_{\rm{mean}}^2=\frac{\sum_{i=1}^N W_i \left( \langle\av\rangle -  \av_{,i}\right)^2 }{(\sum_{i=1}^N W_i)}.
\end{equation}

The signal-to-noise ratio (SNR) of the final map is dependent upon the smoothing technique used.  In order to eliminate suspected foreground stars that are expected to increase the uncertainty in \av~especially in highly extincted regions, we would have liked to smooth the map using sigma-clipping.  In this technique, the local uncertainty in $\langle\av\rangle$, $\sigma_{A_{\rm V}}$, is estimated, and then Equation~\ref{eq:av} is recalculated using only stars that fall within $\pm n$-$\sigma_{A_{\rm V}}$ of the first estimate of \av.

When the distribution of control field colors is sufficiently narrow, or when extinction caused by the clouds is sufficiently high and the fraction of foreground stars low, sigma-clipping is a practical solution to removing foreground stars that contribute to the noise but not the signal.  Otherwise, especially when the latter case is untrue, there is the risk of underestimating the extinction.  This is because sigma-clipping eliminates outliers with both high and low extinction.

\begin{figure}[ht!]
\epsscale{1.}
\plotone{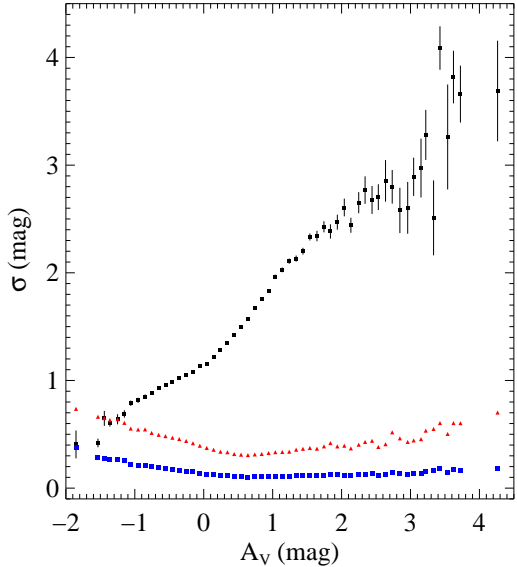}
\caption{The dispersion of measured extinctions as a function of \av~in bins of 0.10 mag.   The black squares show $\sigma_{\rm{mean}}$, as calculated by Equation~\ref{eq:av}, averaged in each bin, with the relative scatters indicated by error bars. For comparison, we plot the mean error due only to scatter in intrinsic color and that due only to photometry, indicated by red and blue triangles, respectively.  \label{sigav}}
\end{figure}

\begin{table}[ht]
\begin{center}
\caption{Summary of LMC reddenings.\label{tab:red}}
\begin{tabular}{ll}
\tableline\tableline
\ebv~ (mag)\tablenotemark{a} & Reference  \\
\tableline
0.12\tablenotemark{b} &  This paper\\
\tableline
0.10  		&1 \\
0.07\tablenotemark{c} 	&2\\
0.02--0.20  	&3 \\
0.10		&4  \\
0.15 (0.01--0.26)\tablenotemark{c}  &5  \\
0.13\tablenotemark{c}  &6\\
0.16   		&7 \\
0.13	   	&8 \\ 
0.13		&9\\
\tableline
\end{tabular}
\tablenotetext{a}{Except for the Isserstedt (1975) value, a maximum \\
detection, values listed are means or ranges of \ebv.}
\tablenotetext{b}{In converting our derived $\langle\av \rangle$ to $\langle E(B-V) \rangle$, we used \\
$R=\av/\ebv=3.1$, as did the authors listed here.  \\
The use of Gordon et al.'s more recent value calculated for \\
the LMC, $R=3.41$, would result in $\langle E(B-V) \rangle=0.11$ mag.}  
\tablenotetext{c}{These authors did not correct for Galactic foreground \\
reddening.}

\tablerefs{(1) Feast et al. (1960); (2) Isserstedt \\
(1975); (3) Isserstedt \& Kohl (1984); (4)  Grieve \\
\& Madore (1986); (5)  Hill et al. (1994); (6)  Massey \\
et al. (1995); (7)  Oestreicher \& Schmidt-Kaler (1996);\\
 (8) Harris,Zaritsky \& Thompson (1997) ; \\
(9) Zaritsky (1999)}
\end{center}
\end{table}

The trend of uncertainty increasing with extinction (see Figure~\ref{sigav}) results from a number of sources of error, including the uncertainty of the intrinsic colors, small-scale structure within pixels, photometric errors in the individual bands, and foreground stars within the LMC.  The latter two uncertainties are expected to increase in highly extincted regions.   The stellar number density in high-\av~pixels is comparable to low-\av~areas of the map ($\sim15$ stars per pixel and $\sim 21$ stars per pixel for $\av>2$ and $\av<2$ mag, respectively), and so Poisson noise is not the main contributor to the increased dispersion.

In the NICER method, $\sigma_i^2$ is calculated for each star from the covariance matrices of the intrinsic scatter in colors and photometric errors.  The corresponding measurement error of a pixel is given by $\sigma_{\rm meas}^2=1/\sum_{i=1}^N W^2_i$.  The following analysis of $\sigma_{\rm meas}^2$ and the total dispersion of a pixel,  $\sigma_{\rm mean}^2$ (Equation~\ref{eq:sig}), can help us determine which types of uncertainties dominate. 

As shown in Figure \ref{sigav}, in low-extinction regions of our map, measurement (i.e., photometric and intrinsic color) errors contribute to a larger fraction of the overall uncertainty than in high-extinction regions.  As the extinction increases, the dispersion due to small-scale structure and foreground stars within the LMC increasingly dominate the uncertainty.  As discussed in previous studies (e.g. Lada et al. 1994; Alves et al. 1998; Lombardi et al. 2006), these two phenomena are difficult to disentangle since they result in similar statistics. For instance, an un-reddened star appearing through a ``hole'' in a GMC is indstinguishable from a foreground star in front of the same cloud.

We check the accuracy of our estimated uncertainties by examining the distribution of pixels with $\av<0$,  The appearance of a significant number ($\sim 1/5$) of such pixels (see Figure~\ref{avhist}) is consistent with a noise population with a standard deviation characterized by the mean measurement error.  That is, stars with $\av<0$ are well-fit by a Gaussian with a standard deviation equal to the propagated measurement uncertainties, $\langle\sigma_{\rm meas}\rangle=0.39$ mag.    Recall, the dispersion of the control group colors---the value that determines the significance level of the extinction measurements---is not much less than the dispersion of the field star colors (see Figure~\ref{control} in \S~\ref{sec:data_reduct}).    With deeper observations and better measurements of the intrinsic colors of a well-defined population, these statistics could be improved upon.   Meanwhile, as indicated by Figure~\ref{avhist}, we are unable to distinguish real signal from noise below $\av\approx 1$ mag.  With an overall 1-$\sigma$ noise level of 0.57, the average SNR of the entire extinction map is 0.66; it is 2.5 for those regions of the map where $\av>1$ mag.

We note that the column densities derived in this paper may be systematically lowered if the extinction coefficients derived in \S~\ref{sec:extlaw} are too large.  This would be the case if we assumed too low of a gas-to-dust-ratio, (see Equation~\ref{eq:extlaw}).

However, the measured columns are more likely to be lower limits, for the following reasons.  (1) Foreground contamination results in the dilution of signal, and thus, lower values of \av. (2) The most highly extincted stars likely went undetected.  In \S~3.3 we show that our measurements of \av~are relatively insensitive where $\ico>2$ \counits; deeper photometry in these regions would raise the measured extinction.

\subsubsection{Comparison with previous studies}

The mean internal extinction we derive here is in overall agreement with earlier measurements.  Table \ref{tab:red} summarizes the mean reddenings given by previous authors, each of whom used some form of the color excess technique---subtracting intrinsic from observed colors---to derive a reddening distribution.  All used $UBV$ photometry of early-type stars for their reddening measurements.  They were necessarily observing the most luminous stars, which may have additional circumstellar reddening due to heavy mass-loss. This could lead to overestimating the interstellar reddening, as Oestreicher \& Schmidt-Kaler (1996) point out.  

Other common sources of error and inaccuracy in the results of previous authors included (i) the assumption of Galactic intrinsic colors, (ii) not accounting for the Galactic foreground contribution to reddening, (iii) not addressing selection effects related to completeness, (iv) small data sets, and (v) limited spatial coverage. Adopting Galactic intrinsic colors (e.g., Feast et al. 1960, Isserstedt 1975, Isserstedt \& Kohl 1984)  leads to an underestimate of the reddening since MW supergiants (the stellar types used by these authors) are intriniscially redder in color than their LMC counterparts, especially in $UBV$ photometry.  Not accounting for the completeness limit (e.g., Massey et al. 1995) also biases the measurements toward lower extinction, since highly reddened faint stars are underrepresented (see \S~2.2.1).  The extinction will be {\em overestimated}, however, if the Galactic foreground contribution to reddening isn't corrected (e.g.,  Hill et al., Massey et al. 1995).

Inaccuracies in the overall spatial distribution of reddening may result from items (iv) and (v).  Feast et al. (1960) had a small data set (108 supergiants) sampled from across the LMC.  Hill et al. (1994) used 7 OB associations, 3 of which are associated with 30 Doradus, one of the most highly reddened regions of the galaxy.  Harris et al. (1997) and Zaritsky (1999), as part of connected studies, produced maps of variable spatial resolution which covered areas of $\sim 2^\circ\times 2^\circ$ and  $\sim 4^\circ\times 2.7^\circ$, respectively.  The most comparable map to ours is that created by Oestreicher \& Schmidt-Kaler (1996) from 1,507 O-A stars throughout the LMC.  They did not grid their map, but rather analyzed the distribution of reddened stars.  Like us, they found reddening throughout the LMC, including in peripheral areas not associated with \HI~or CO peaks. They also did not find much reddening in the bar.


\subsection{The X-Factor}\label{sec:xfactor}

\begin{figure*}[ht!]
\epsscale{1.9}
\plotone{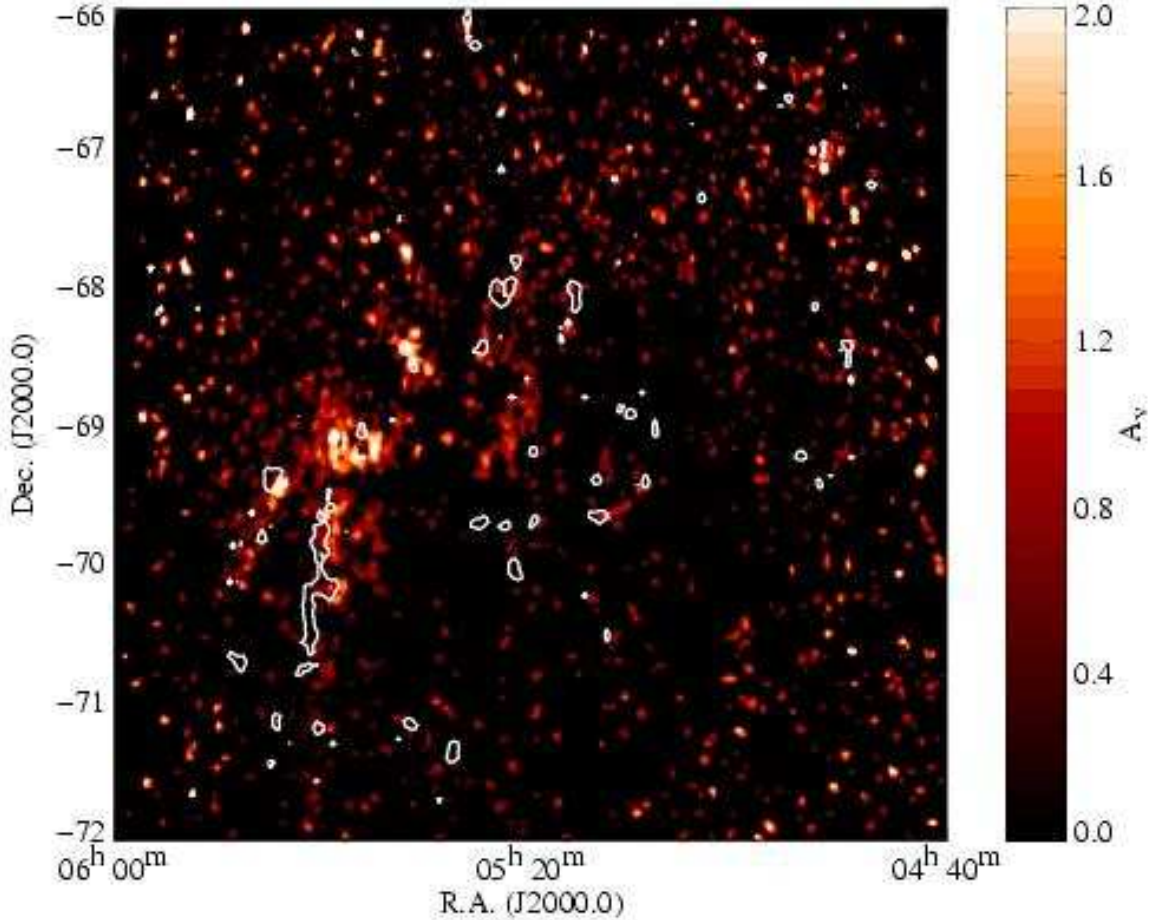}
\caption{This map shows extinction due to H$_2$.  It is created by applying Equation~\ref{nh2} to the NICER extinction map (Figure~\ref{avmap1}), using only pixels with total $\av>1$ mag.  Overlaid are the CO contours of molecular clouds identified by Mizuno et al. \cite{mizuno}.\label{avmap2}}
\end{figure*}

\begin{figure}[h]
\epsscale{.95}
\plotone{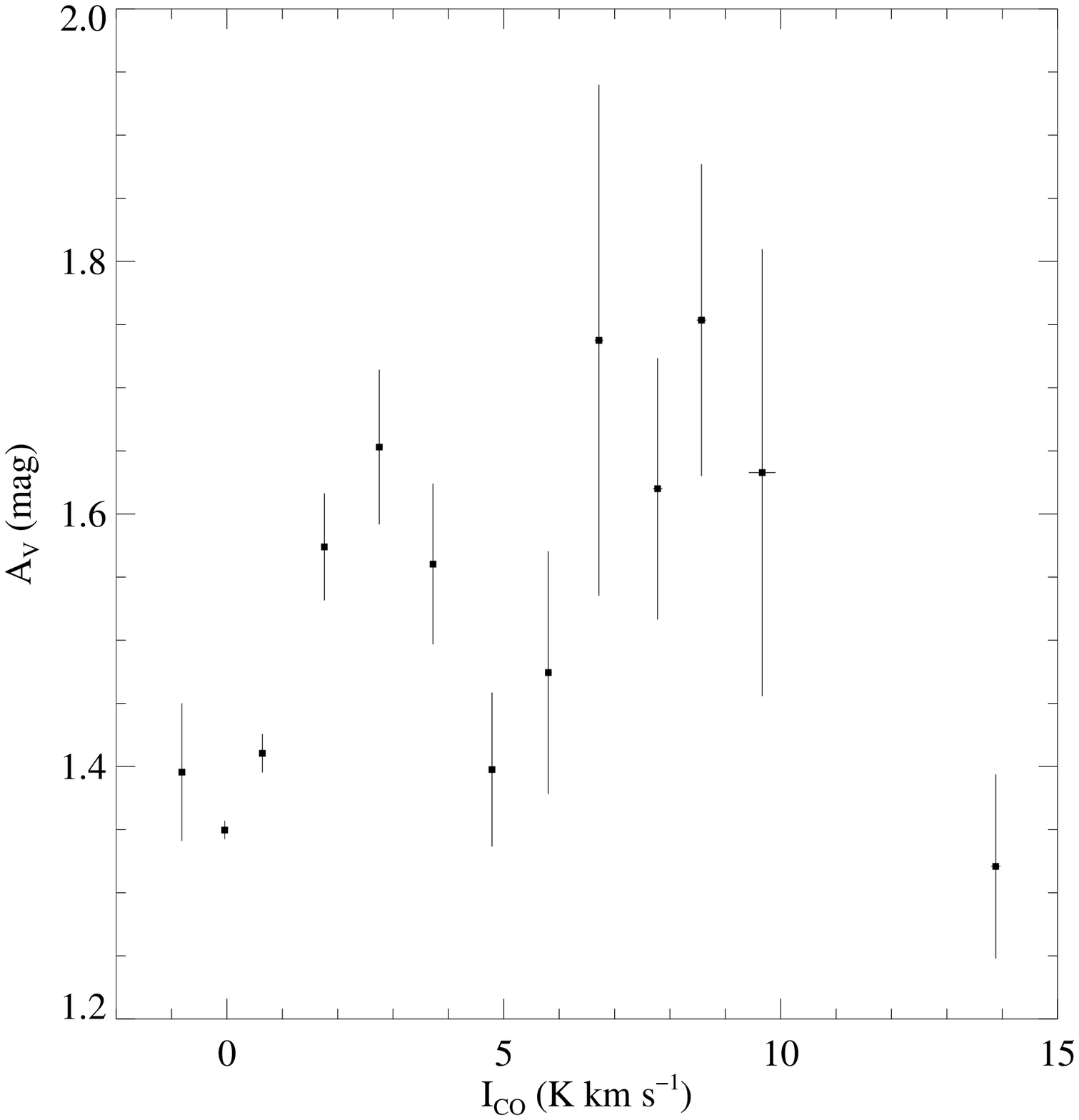}
\caption{Plotted is the \ico-\av relation for the analyzed regions (see the text for details).  Data are binned along the \ico~axis in bins of $\Delta\ico=1.0$ \counits.   \label{xfac}}
\end{figure}

The $X$-factor is defined as
\begin{equation}\label{eq:xfac}
X\equiv \frac{N(\H_2)~ [\cm]}{\ico~[\counits]}.
\end{equation}

To determine \x{LMC}, both the CO data and NICER extinction map were convolved to maps of the same pixel size ($4'$) and grid spacing ($2'$) and then compared on a pixel-by-pixel basis. These maps are displayed in Figure~\ref{avmap2}. For the reasons discussed in the previous section, only pixels in which $\av\ge 1$ mag (or, SNR$\ge 2.5$) are used. A linear relationship between $\av$ and $\ico$ would allow us to derive $X$ by fitting the data. However, above about 2 \counits, as shown in Figure~\ref{xfac}, there does not appear to be any linear dependence between \ico~and \av. This lack of a linear correlation at high column densities most likely results from insufficient sensitivity in our extinction measurements. The global mean SNR of the \av~map (0.66) surpasses that of the \ico~map (0.23). But the comparative sensitivities of the two measurements appear to depend on the molecular content in a given area.  In high column density regions, above $\ico=2~\counits$, these SNRs are 1.60 for \av~and 6.5 for \ico.  In low column density regions, $\ico<2~\counits$, the trend is reversed and the ratios are 0.82 and 0.18, respectively.

In order to take full advantage of the data in the regions of most concern to us, i.e., regions of highest extinction, we proceed by deriving the $X$-factor statistically.  Of the 4,396 pixels with $\av\ge 1$ mag, there are 271 remaining corresponding to positions where $\ico\ge 2$ \counits.  

The total contribution of hydrogen gas to the column density, $N(\H_{\t{tot}})\equiv N(\HI)+2N(\H_2)$, may be obtained from \av.  Statistically, along any given line-of-sight toward the LMC, half the total column contributes to the measured extinction. That is, $N(\H_{\rm tot}^{(\av)})=N(\H_{\rm tot})/2$. In terms of the average gas-to-dust ratio calculated by Gordon et al. \cite{gordon}, the column density of molecular hydrogen is

\begin{eqnarray}
N(\H_2)[\rm{cm}^{-2}]&=&3.25\times 10^{21}\av[\rm mag] \nonumber \\
 &-&\frac{1}{2}N(\HI)[\rm{cm}^{-2}]. \label{nh2}
\end{eqnarray}

\noindent Figure~\ref{avmap2} displays the map of visual extinction due to $\H_2$, created using this equation.  Finally, the mean of Equation~\ref{eq:xfac} is calculated for pixels where $\av\ge 1$ mag and $\ico \ge 2$ \counits.  We note that the suitability of this method relies on $2N(\H_2)$ being significantly larger than $N(\HI)$ since, otherwise, large errors would result from the subtraction in Equation~\ref{nh2}.  For our measurements towards regions of the LMC where $\av\ge 1$ mag, $2N(\H_2)/N(\H_{\rm tot})=73\%\pm 17\%$, indicating that it is reasonable to proceed with  Equation~\ref{nh2}.  

In units of $10^{20}~\xunits$, we found that $\x{LMC}=9.3\pm 0.4$, where the error around the mean (i.e., the random error) was adopted as the uncertainty.  Values range from 0.10 to 47.1, and the 1-$\sigma$ scatter is 6.9. Figure \ref{xhist} displays this distribution.  In the same regions, we measure a total $\H_2$ mass of $(4.5\pm 0.2)\times 10^7\msun$ assuming a distance to the LMC of 50 kpc.

\begin{figure}[h]
\epsscale{1.}
\plotone{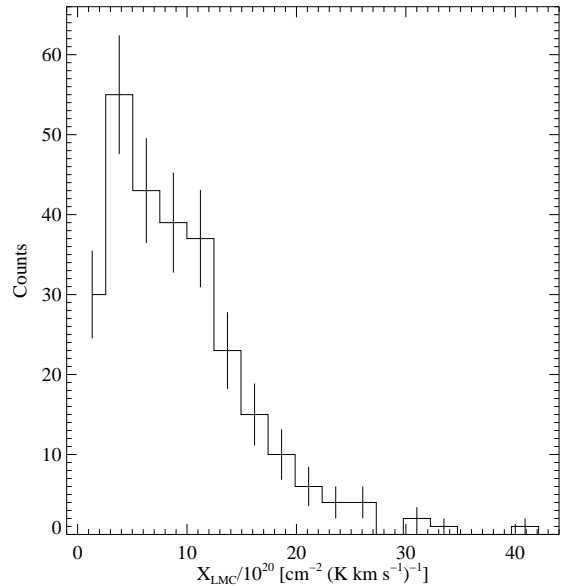}
\caption{The distribution of \x{LMC}~for the analyzed regions.  The error bars are the deviation in the mean number of counts for each bin. \label{xhist}}
\end{figure}

\subsubsection{Comparison with previous studies}

 Our value of \x{LMC}, $\sim 4.7\x{MW}$, falls in between that of Mizuno et al. (2001) ($\sim 4.5\x{MW}$), who calculated it from high resolution ($2.'6$) CO observations, and that of Israel \cite{israel} ($\sim 6.5\x{MW}$; see Table \ref{tab:xfac}). The former group  assumed that CO clouds in the LMC are virialized and calculated the $X$-factor using $X=M_{\rm vir}/L_{\rm CO}$, where $M_{\rm vir}$ is the virial mass of a cloud and $L_{\rm CO}$ is the CO luminosity.   Application of the virial theorem to complex molecular structures, known to exist in the LMC, may be inappropriate.  There is no {\em a priori} reason why the elongated structures associated with 30 Doradus and the supershell regions, for instance, should be virialized. Our value is also significantly larger than the Blitz et al. (2006), who did a revised calculation based on the Mizuno et al. (2001) observations, of $\sim 2.7\x{MW}$.  If the GMCs were not virialized, we would have expected in our calculations a smaller value of \x{LMC} than that obtained by Blitz et al. (2006).  Still, their definition of what constitutes a molecular cloud may have led to an underestimate in the total molecular mass.

\begin{table}[h]
\begin{center}
\caption{Summary of LMC $X$-factors.\label{tab:xfac}}
\begin{tabular}{ll}
\tableline\tableline
$\x{LMC}/10^{20}~\xunits$   &   Reference\\
\tableline
$9.3\pm 0.4$		&This paper \\
\tableline
 17  			&1   \\
39 \tablenotemark{a}   	&2	\\
$13\pm 2$    		&3    \\
$9\pm 4$		&4	\\
$5.4\pm0.5$ 		&5	\\
\tableline
\end{tabular}
\tablenotetext{a}{This value was calcluated for the 30 Doradus region. \\
 All others are global LMC averages.}
\tablerefs{(1) Cohen et al. \cite{cohen}; \\
(2) Garay et al. \cite{garay}; (3) Israel \cite{israel}; \\
(4) Mizuno et al. \cite{mizuno}; (5) Blitz et al. \cite{blitz}}
\end{center} 
\end{table}

However, the molecular mass may be overestimated when, at limited resolution, unrelated clouds at different velocities are blended within a beam.   At their resolution ($8.'8$), Cohen et al. \cite{cohen} were able to resolve only the largest cloud complexes ($\sim 140$ pc).  They derived a value of \x{LMC} nearly 2 times our own, though their results did not assume virialization.  Finding that LMC clouds are roughly 6 times fainter in CO than Galactic clouds, Cohen et al. \cite{cohen} reasoned that $\x{LMC}\approx 6 \x{MW}$, where they adopted Bloemen et al.'s \cite{bloemen86} value of  $\x{MW}=2.8$.  But using the updated value of $\x{MW}=2.0$ (Bloemen 1989), would result in a value ($\x{LMC}=12.1$) a bit closer to ours.

Garay et al. \cite{garay}, using the same logic and value of \x{MW}~as Cohen, found that $\x{LMC}\approx 39$ in the region of 30 Doradus.  Israel \cite{israel} found an even greater value (84), while Mizuno et al. \cite{mizuno} observed no significant difference in the $X$-factor there compared to other regions.  The highest value we measure in this region is 33.7, located at $(\alpha,\delta)=(5^{\rm h}43^{\rm m}51^{\rm s}, -69^\circ 26^{\rm m})$.  Our relatively low values in this region may result from underestimating $N(\H_2)$ (see \S ~\ref{sec:xfactor}).  The largest value we findoverall, 47.1, is located at $(5^{\rm h}35^{\rm m}05^{\rm s}, -67^{\rm d}38^{\rm m})$ and is associated with a cloud identified by Mizuno et al. \cite{mizuno} as having an integrated intensity of 4.2 \counits.

Deriving the $X$-factor independently of virialization assumptions by calculating $N(\H_2)$ from a comparison of far-IR surface brightness and $N(\HI)$, Israel \cite{israel} also measured a larger value ($\sim 6.5\x{MW}$) than our own.  He gives a number of reasons for why his measurements of  $N(\H_2)$, and therefore \x{LMC}, are lower limits.  But it appears that his \ico~measurements, borrowed from Cohen et al. \cite{cohen}, are the reason for his comparatively large value of \x{LMC}, since $\x{}\propto\ico^{-1}$.  The mean integrated intensity of the 22 clouds Israel \cite{israel} uses to determine \x{LMC}~is 1.40 \counits; the mean \ico~ detected in corresponding positions by Mizuno et al. \cite{mizuno} is $\sim 8.5$ \counits.  This dimunition in measured CO flux by Israel \cite{israel} was likely a result of his convolving Cohen et al.'s \cite{cohen} original map to a coarser resolution of $15'$.

Previous authors' derivations of \x{LMC}, whether over- or underestimated, necessarily tie to their measurements of total $\H_2$ mass.  Cohen et al. \cite{cohen}, Israel \cite{israel}, and Mizuno et al. \cite{mizuno} derive values of $1.4\times 10^8~\msun$, $(1.0\pm 0.3)\times 10^8~\msun$, and 4--7$\times 10^7~\msun$, respectively.  Our value, $(4.5\pm 0.2)\times 10^7~\msun$, as well as our $\H_2$-to-\HI~mass ratios are consistent with Mizuno et al. \cite{mizuno}.  Based on the \HI~data of Staveley-Smith et al. (2003), we measure a global ratio of $\sim 10\%$. In high-column density regions, where $N(\HI)> 1.5\times 10^{21}~\cm$ as defined by Mizuno et al. \cite{mizuno}, we find a ratio of $\sim 50\%$.  This consistency in mean global measurements is probably due to our relative sensitivities: the CO observations are more sensitive in high-column (i.e., high-mass) regions, whereas our NICER map is more sensitive in the more numerous, low-column regions of the LMC.  

Further in keeping with the relative sensitivities of the NICER technique versus CO observations, discussed in \S~3.3, is the appearance of our maps.  With the exception of the 30 Doradus region, there are few coincidences between the peaks of our $\H_2$ extinction map and GMCs identified by Mizuno et al. \cite{mizuno}, as shown in Figure \ref{avmap2}.  The group catalogued a total of 107 molecular clouds, including 55 ``large'' molecular clouds.  But throughout the LMC, we measure many places where the $\H_2$ is not associated with CO.  In a few regions, exemplified by the areas $(\alpha,\delta)\sim(5^{\rm h}40^{\rm m}, -69^\circ)$ and $\sim(5^{\rm h}30^{\rm m}, -68^\circ 30')$, $\H_2$ and CO overlap, with the former extending further than the latter.  More commonly, CO and $\H_2$ regions appear adjacent to one another.  This suggests that apparently discrete structures are sometimes part of larger complexes that, observed with either method alone, cannot be seen in their entirety.  For instance, the three clumps extending from  $\sim(5^{\rm h}25^{\rm m}, -66^\circ)$---an $\H_2$ peak bordered by two CO GMCs---may be part of one, elongated structure with varying column density.


\section{Conclusions}
We have used the NICE (Lada et al. 1994) and NICER (Lombardi \& Alves 2001) techniques to derive, respectively, the NIR reddening law and \av~distribution of the LMC.  Despite the large number of assumptions and uncertainties necessarily introduced in applying these methods to extragalactic sources, results were largely consistent with previous studies.  Moreover, with the unprecedentedly large data set from 2MASS, the measured \av~distribution may indeed be an improvement. Our main results are as follows:

\begin{enumerate}
\item
By comparing the \HI~and \ex~distributions (the latter derived with the NICE method) on a pixel-by-pixel basis, we calculated the NIR extinction coefficients for the LMC to be $\av/\ehk=20.83\pm0.52$ and $\av/\ejh=17.30\pm 0.46$. Dividing these results $\ejh/\ehk=1.20\pm0.04$ gives a result consistent with independently measured reddening laws.

\item
Using the NICER technique, we created an \av~map whose spatial distribution showed many areas of high column density gas across the face of the LMC that were traced by little CO emission. The mean value of 0.38 mag is consistent with previous results.
  
\item
We created an $\av(\H_2)$ map and compared it to the \ico~map provided by Mizuno et al. \cite{mizuno}.  In regions of significant CO emission we found that $X$-factor of $(9.3\pm 0.4)\times 10^{20}~\xunits$ and $M(\H_2)=(4.5\pm0.2)\times 10^7~\msun$.  Our value of $\x{LMC}$ is less than some authors who, using other methods, probably overestimated it.  Since our value is greater than that calculated by both Mizuno et al. (2001) and Blitz et al. \cite{blitz}, this is consistent with their assumption the LMC GMCs are virialized.

\end{enumerate}

\begin{acknowledgements}
We thank Lister Staveley-Smith for providing us with the \HI~maps of the Galactic foreground and LMC.  We also thank Norikazu Mizuno and Yasuo Fukui for providing the CO data.

\end{acknowledgements}

\end{document}